\documentclass[aps,prl,twocolumn,showpacs,amsmath,amssymb,amsfonts,nofootinbib,long]{revtex4}
\usepackage{epsfig,latexsym,bm}

\newcommand\MeV{\mbox{MeV}}
\newcommand\GeV{\mbox{GeV}}

\newcommand\kpc{\mbox{kpc}}
\newcommand\Mpc{\mbox{Mpc}}

\newcommand\G{\mbox{G}}
\newcommand\nG{\mbox{nG}}

\newcommand\B{\mathbf{B}}

\newcommand\x{\mathbf{x}}

\newcommand\mPl{m_{\rm Pl}}

\newcommand\dec{\mbox{\scriptsize dec}}

\begin{document}

\title{A MODEL OF UNIVERSE ANISOTROPIZATION}

\author{Leonardo Campanelli$^{1,2}$}
\email{leonardo.campanelli@ba.infn.it}

\affiliation{$^1$Dipartimento di Fisica, Universit\`{a} di Bari, I-70126 Bari, Italy}
\affiliation{$^2$INFN - Sezione di Bari, I-70126 Bari, Italy}

\date{September, 2009}


\begin{abstract}
\begin{center}
{\bf Abstract}
\end{center}

The presence of a nonconformally invariant term in the photon sector
of the Lorentz-violating extension of the standard model of
particle physics, the ``Kosteleck\'{y} term''
$\mathcal{L}_K \propto (k_F)_{\alpha \beta \mu \nu} F^{\alpha \beta} F^{\mu \nu}$,
enables a superadiabatic amplification of magnetic vacuum
fluctuations during de Sitter inflation. For a particular form of the external
tensor $k_F$ that parametrizes Lorentz
violation, the generated field possesses a planar symmetry at large
cosmological scales and can have today an intensity of order of
nanogauss for a wide range of values of parameters defining
inflation. This peculiar magnetic field could account for the
presently observed galactic magnetic fields and
induces a small anisotropization of the Universe at cosmological scales. 
The resulting Bianchi I model could explain the presumedly low quadrupole
power 
in the cosmic microwave background radiation.

\end{abstract}


\maketitle

\newpage


\section{\normalsize{I. Introduction}}
\renewcommand{\thesection}{\arabic{section}}


The standard model of particle physics together with the Einstein theory of gravity
encompasses, basically, all known fundamental physics. However, due to the classical
character of general relativity, these theories are believed to be incomplete,
and a search for a more fundamental theory which overcomes the stumbling block
of quantizing gravity is a major goal of present-day theoretical physics.
A promising candidate is string theory, which deals with gravity in a quantum
and self-consistent way. Although string theory is well away from being an
experimentally tested theory, some of its peculiar, low-energy manifestations, if
ever detected in astrophysical and/or ground-based experiments,
could be a strong signal of its correctness. Indeed, as pointed out by
Kosteleck\'{y}~\cite{Kostelecky0}, a possible detectable signature
of string theory at low energies is violation of Lorentz symmetry~\cite{Kostelecky}.


If on the one hand, terrestrial and astrophysical experiments
have not yet either confirmed or ruled out the existence of effects of
Lorentz violation (LV), on the other hand
astrophysical observations have definitely confirmed the presence of large-scale
correlated, microgauss magnetic fields in any type of galaxies and clusters of galaxies.
(For reviews on cosmic magnetic fields see Ref.~\cite{Widrow}.)
This peculiar property of possessing, roughly speaking,
the same intensity and correlation scale everywhere
in the present Universe, has been interpreted as a strong hint that
cosmic magnetic fields are, indeed, relic of the early Universe.
Obviously, a damning evidence for the primordial origin of cosmic
magnetic fields would be the detection of their effects on the
Cosmic Microwave Background (CMB) radiation. Until now, however,
exhaustive analyses have only given stringent constraints on their properties,
without finding any evidence of their imprints on the CMB~\cite{Dolgov}.


The use of the CMB radiation as a probe of the physics of the early Universe
has proved to be very fruitful in the last years. In particular,
the high resolution data of temperature fluctuations of CMB angular power spectrum,
provided by the Wilkinson Microwave Anisotropy Probe (WMAP)~\cite{WMAPsite,WMAP5} have,
almost definitively, consecrated the so-called ``$\Lambda$-dominated cold
dark matter'' ($\Lambda$CDM) as the standard cosmological model of the Universe.

Nevertheless, the 1-, 3-, and 5-year WMAP data display at large angular
scales some ``anomalous'' features, the most important ones being
the low quadrupole moment and the presence of a preferred direction in the Universe.

It is extremely important, however, to stress that, due to the importance of
residual Galactic foreground emission, those
anomalies in the CMB anisotropy are still subject to an intense
debate~\cite{AnomalyDebate}.

If there is really a problem with the quadrupole moment,
then its lowness, indicating a suppression of power
at cosmological scales, may signal a nontrivial topology of
the large-scale geometry of the Universe~\cite{Topology}. Indeed,
several possibilities have been proposed in the recent literature to
understand that suppression~\cite{Quadrupole1,Quadrupole2} (for other large-scale
anomalies in the angular distribution of CMB see, e.g.,
Ref.~\cite{Anomalies}). Recently enough~\cite{prl,prd},
it has been shown that a particular case of the simplest anisotropic cosmological model,
i.e. the Bianchi I model, could account for the smallness of the quadrupole,
without affecting higher multipoles of the angular power spectrum of
the temperature anisotropy. Such a proposal of an ``ellipsoidal universe''
has been considered also in Ref.~\cite{Gruppuso,Ge}.

Also, the WMAP data display a particular feature which has been deeply investigated
in the last years: a statistically significant alignment and planarity of the
quadrupole and octupole modes. This seems to indicate the existence
of a preferred direction in the Universe, which has been named ``axis of
evil'' (AE)~\cite{Land-Magueijo}. Needless to say,
there is no space in the isotropic, standard cosmological model for such a type of features.


In this paper, we investigate the possibility that effects of Lorentz violation
at inflation could be responsible for the creation of large-scale
magnetic fields possessing planar symmetry. As we will see, these fields can have the
right intensity and correlation length to explain the existence of
galactic and extragalactic magnetic fields.
Moreover, because of their peculiar symmetry at cosmological scales,
they could induce a modification of the (isotropic) Robertson-Walker metric,
in such a way that the resulting cosmological model is well described by
the ellipsoidal universe model. This, in turn, could naturally account for some
peculiar and not-yet-explained anomalous features of CMB radiation discussed above.

The plan of the paper is as follows. In Sec. II we discuss the
generation at inflation of a plane-symmetric cosmic magnetic field
in the framework of the Lorentz-violation extension of the standard
model of particle physics. Section III deals with the analysis of CMB
anisotropies including the asymmetric contribution due to the
presence of the planar field.
Finally, we draw our conclusions in Sec. IV. Some technical details
are presented in the appendixes.


\section{\normalsize{II. Lorentz-violating Electromagnetism
and Planar Cosmic Magnetic Fields}}
\renewcommand{\thesection}{\arabic{section}}

The large correlation scale of cosmic magnetic fields,
ranging from $\sim 10 \kpc$ for magnetic fields in
galaxies to $\sim 1\Mpc$ for those in clusters, and the fact that they
are found to have approximately the same intensity of a few microgauss
seems to indicate a common and primordial origin, probably to ascribe to
some unknown mechanism acting during an inflationary epoch of the Universe.
If one takes into account that the collapse of primordial large-scale structures
enhances the intensity of any preexisting magnetic field of about a
factor $10^3$~\cite{Widrow}, a primeval field with comoving
intensity of order of nanogauss and correlated on megaparsec scales
could explain the ``magnetization of the Universe''.

During inflation all fields are quantum mechanically excited.
Because the wavelength $\lambda$ associated to a given fluctuation grows faster
than the horizon, there will be a time, say $t_1$, when this mode crosses
outside the horizon itself. After that, this fluctuation cannot
collapse back into the vacuum being not causally self-correlated,
and then ``survives'' as a classical real object~\cite{Dimopoulos}.

The electromagnetic energy density at the time of crossing is then
fixed by the Gibbons-Hawking temperature $T_{\rm GH}$~\cite{Dimopoulos}:
\begin{equation}
\mathcal{E} \sim T_{\rm GH}^4 \sim H^4,
\end{equation}
where $H$ is the Hubble parameter (in this paper we consider, for the
sake of simplicity, just the case of de Sitter inflation).
Taking into account the expression for the electromagnetic energy in
standard Maxwell electromagnetism, one
arrives to the result that the spectrum of magnetic fluctuations at
the time of horizon crossing is given by
$B_1 \sim H^2 \sim M^4/\mPl^2$~\cite{Dimopoulos,Turner,nonlinear},
where in the last equality we used the Friedmann equation $H^2 = (8\pi/3) M^4/\mPl^2$.
Here, $M^4$ is the total energy density during inflation (which is
constant during de Sitter inflation) and $\mPl \sim 10^{19} \GeV$ is
the Planck mass. Because of conformal invariance of Maxwell
electromagnetism one finds, however, that the present magnitude of
the inflation-produced field at the scale, say $10 \kpc$, is
vanishingly small, $B_0 \sim 10^{-52} \G$~\cite{Turner}.
(This is true only if the background metric is
spatially-flat~\cite{Barrow}, which is the case discussed in this
paper.) Since the pioneer work of Turner and Widrow~\cite{Turner}, a
plethora of mechanisms has been proposed for generating cosmic
magnetic fields in the early Universe, all of which repose on the
breaking of conformal invariance of standard electrodynamics
(see references in Ref.~\cite{Widrow} and, for recent papers,
Ref.~\cite{Generation,Dimopoulos2}).

In particular, Kosteleck\'{y}, Potting and Samuel~\cite{Kostelecky-p}
first pointed out that the breaking of
conformal invariance is a natural consequence of LV. Indeed, they
argued that the appearance of an effective photon mass, owing to
spontaneous breaking of Lorentz invariance, could enable the
generation of large-scale magnetic fields within inflationary
scenarios. The idea that Lorentz symmetry breaking could result in the
generation of cosmic magnetic fields has been pursued since
then by others authors~\cite{Bertolami,Mazumdar,Bamba,Gamboa,Campanelli}.


The aim of this paper is to show that within a particular Lorentz-violating model of
particle physics, it is possible to generate magnetic fields of cosmological type
possessing a peculiar spatial geometry.
Then, in the next section, we will analyze their impact on the isotropy of the Universe
and, in particular, on the cosmic microwave background radiation.

The model we are going to study is the so-called
standard model extension (SME)~\cite{Colladay}, which is
an effective field theory including all admissible
Lorentz-violating terms in the Glashow-Weinberg-Salam gauge theory.
In curved spacetimes, the SME action for the photon field, here referred to as the
Maxwell-Kosteleck\'{y} (MK) action, reads~\cite{Kostelecky1}
\begin{equation}
\label{Action} S_{\rm MK} = \int \!\! d^4x \, e \! \left[ -
\mbox{$\frac14$} \, F_{\mu \nu} F^{\mu \nu} - \mbox{$\frac14$} \,
(k_F)_{\alpha \beta \mu \nu} F^{\alpha \beta} F^{\mu \nu} \right]
\!,
\end{equation}
where $F_{\mu \nu} = \partial_{\mu} A_{\nu} - \partial_{\nu}
A_{\mu}$ is the electromagnetic field strength tensor and $e$ the
determinant of the vierbein. The presence of the external tensor
$(k_F)_{\alpha \beta \mu \nu}$ breaks (particle)
Lorentz invariance~\cite{Kostelecky1} and parametrizes then Lorentz violation.
The external tensor $(k_F)_{\alpha \beta \mu \nu}$ is fixed in a given system of
coordinates. Going in different systems of coordinates will,
generally, induces a change of the form of $(k_F)_{\alpha \beta
\mu \nu}$. In the following, we assume
that the form of $(k_F)_{\alpha \beta \mu \nu}$ refers to a system
of coordinates at rest with respect to the cosmic microwave
background, the so-called ``CMB frame.''

It is worth noting that taking $(k_F)_{\alpha \beta \mu \nu}$ as a
fixed tensor corresponds to have an explicit violation of Lorentz
symmetry. This could introduce in the theory an instability
associated to nonpositivity of the energy. Indeed,
working in a flat isotropic universe described by a Robertson-Walker
metric $ds^2 = a^2(d\eta^2 - d \x^2)$, where $a(\eta)$ is the
expansion parameter and $\eta$ the conformal time, and introducing the electric and
magnetic fields as $F_{0i} = -a^2 E_i$ and $F_{ij} =
\epsilon_{ijk} a^2 B_k$ (Latin indices run from $1$ to $3$, while
Greek ones from $0$ to $3$), the electromagnetic energy density
turns out to be~\cite{Campanelli-Cea}
\begin{eqnarray}
\label{Energy} \mathcal{E} \!\!& = &\!\! 
\frac12 ({\textbf E}^2 + {\textbf B}^2) + (k_F)_{i00j} a^{-4} E_i E_j \nonumber \\
\!\!& + &\!\! \frac14 (k_F)_{ijkl} a^{-4} \epsilon_{ijm} \epsilon_{kln} B_m B_n.
\end{eqnarray}
The positivity of the above quadratic form depends on the particular
form assumed by the fixed tensor $(k_F)_{\alpha \beta
\mu \nu}$, which is frame-dependent. This apparent paradox
(frame-dependent positivity of the energy)
is overcome when considering models in which LV is
spontaneously broken. In this case, however, action~(\ref{Action}) loses
its character of generality, since the tensor
$(k_F)_{\alpha \beta \mu \nu}$ is now regarded as a vacuum
expectation value of some tensor field with its own dynamics. For
this reason, and following Ref.~\cite{Campanelli-Cea}, we will take
$(k_F)_{\alpha \beta \mu \nu}$ to be a
fixed tensor but, at the same time, we will impose positivity of the
energy in the CMB frame. Needless to say, if in another system of coordinates
the energy is not positive defined, this
means that the effective theory with explicit Lorentz
symmetry breaking becomes meaningless in that frame and
one needs to consider the full theory with spontaneous Lorentz
symmetry breaking in order to get physically acceptable results.



The equations of motion follow from action~(\ref{Action})~\cite{Campanelli-Cea}:
\begin{eqnarray}
\label{motion} && \partial_\eta (a^2 E_i) - \epsilon_{ijk}
\partial_j (a^2 B_k) + \nonumber \\
&& \partial_\eta [2 (k_F)_{i00j} a^{-2} E_j -
(k_F)_{0ijk} \epsilon_{jkl} a^{-2} B_l] + \nonumber \\
&& \partial_j [2 (k_F)_{ijk0} a^{-2} E_k - (k_F)_{ijkl}
\epsilon_{klm} a^{-2} B_m] = 0
\end{eqnarray}
and
%
%
$\partial_i (a^2 E_i) + \partial_i[ 2 (k_F)_{i00j}
a^{-2} E_j - (k_F)_{0ijk} \epsilon_{jkl} a^{-2} B_l] = 0$.
The Bianchi identities are
%
%
$\partial_\eta(a^2 {\textbf B}) + \nabla \times
(a^2 {\textbf E}) = 0\,$ and $\, \nabla \cdot
{\textbf B} = 0$.
%
%
%
%


We are interested in the generation and evolution of
superhorizon magnetic fields, that is to electromagnetic modes
whose physical wavelength is much greater than the Hubble radius
$H^{-1}$, $\lambda_{\rm phys} \gg H^{-1}$, where $\lambda_{\rm phys}
= a \lambda$ and $\lambda$ is the comoving wavelength. Since $a \eta
\sim H^{-1}$, introducing the comoving wavenumber $k =
2\pi/\lambda$, the above condition reads $|k\eta| \ll 1$. Observing
that the first Bianchi identity gives on large scales $B \sim k\eta
E$, where $B$ and $E$ stand for the average magnitude of the
magnetic and electric field intensities, and assuming that all
(non-null) components of $(k_F)_{\alpha \beta \mu \nu}$ have
approximately the same magnitude, we can neglect in
Eq.~(\ref{motion}), on large scales, the terms proportional to the magnetic field.
Also, the next-to-last term is negligible with respect to the first one. Therefore, at
large scales, Eq.~(\ref{motion}) reduces to
\begin{equation}
\label{evolution1} \partial_\eta (a^2 E_i) +
\partial_\eta [2 (k_F)_{i00j} a^{-2} E_j] = 0.
\end{equation}
Assuming that $||(k_F)_{i00j}|| \gg a^4$ we then have
\begin{equation}
\label{evolution2} (k_F)_{i00j} a^{-2} E_j = c_i,
\end{equation}
where $c_i$ are constants of integration.

It is plausible to assume that the tensor $(k_F)_{\alpha \beta \mu
\nu}$ is a nonincreasing function of time. Indeed, we can take the
simple form $(k_F)_{i00j} \propto a^{-p}$, with $p$ a non-negative
real number. In this case, it is clear from Eq.~(\ref{evolution2}),
the first Bianchi identity, and the fact that $\eta \propto a^{-1}$
during de Sitter inflation, that the average intensity of the
magnetic field grows ``superadiabatically,'' $B \propto a^{1+p}$.
[If the elements of $(k_F)_{i00j}$ are either all zero or negligibly
small in the CMB frame, the inflation-produced field is, in general,
small-scale correlated and then not astrophysically interesting.]

In Ref.~\cite{Campanelli-Cea}, the case was analyzed where
$(k_F)_{i00j}$ is a constant isotropic tensor, $(k_F)_{i00j} \propto
\delta_{ij}$, where $\delta_{ij}$ is the Kronecker delta.
In this paper, instead, we study the case in which $(k_F)_{i00j}$ is
``maximally'' anisotropic, in the sense that only one component, say
$(k_F)_{3003}$, is different from zero.
In this case, only the $E_z$ component of the electric field is amplified
and this in turn means that only the $B_x$ and $B_y$ components of the magnetic
field grow superadiabatically during inflation. Accordingly, the amplified
magnetic field will possess a planar symmetry, whose plane of
symmetry is the $xy$-plane.

Moreover, for the sake of simplicity, we will concentrate only on a
single case, that corresponding to $p=2$, so that we will assume that
\begin{equation}
\label{k3003} (k_F)_{i00j} \equiv a^{-2} \delta_{i3} \delta_{j3} k_F,
\end{equation}
where $k_F$ is a constant which gives the magnitude of Lorentz violation
effect at the actual time, $a = a_0 = 1$. (It is worth noting that the condition
$||(k_F)_{i00j}|| \gg a^4$ translates into $k_F \gg a^6$.)
As we will see, the choice of Eq.~(\ref{k3003})
will correspond to the case of inflation-produced, scale-invariant,
cosmic magnetic fields.
\footnote{Up to today there is no experimental evidence that the tensor
$(k_F)_{\alpha \beta \mu \nu}$ is time-dependent or even different from
zero (see below). Accordingly, it is worth stressing that our assumption
that $(k_F)_{\alpha \beta \mu \nu}$ decreases in time should only be considered
as a working hypothesis. On the other hand, the choice $p=2$ follows by the
observation that the actual cosmic magnetic fields possess,
approximately, the same intensity on different scales (that of galaxies
and that of clusters of galaxies), and by interpreting this occurrence as
a hint that their spectrum is indeed scale-invariant
(at least on cosmological scales).}
In this case, the electric and magnetic
fields scale as $E_z \propto a^4$ and $B_x \propto B_y \propto a^3$,
respectively, while $E_x$, $E_y$, and $B_z$ scale adiabatically.


Before proceeding further, we estimate the spectrum of magnetic
vacuum fluctuations generated during de Sitter inflation in
Maxwell-Kosteleck\'{y} electromagnetism. If, as before, we assume that the only nonzero component of
$(k_F)_{\alpha \beta \mu \nu}$ is just given by $(k_F)_{3003} = k_F a^{-2}$, the electromagnetic
energy density turns out to be $\mathcal{E} = \frac12 ({\textbf E}^2 +
{\textbf B}^2) + k_F a^{-6} E_z^2$. Therefore, at the time of
crossing, where $|k\eta| \sim 1$, and since $k_F \gg a^6$, we
get $\mathcal{E} \sim k_F a_1^{-6} B_1^2$, where $a_1 = a(t_1)$ and
we used the first Bianchi identity. Here, $B_1$ stands for the
average intensity of the magnetic field on the plane $xy$ at the
time of crossing. Remembering that $\mathcal{E} \sim H^4$, we obtain
the spectrum of magnetic fluctuations when crossing the horizon,
\begin{equation}
\label{B1} B_1 \sim \frac{a_1^3 H^2}{\sqrt{k_F}} \, .
\end{equation}
%
It is worth noting that, in order
to have positivity of the energy, we are forced to assume $k_F > 0$.


In order to find the actual value of the inflation-produced magnetic field,
we have to follow its evolution during reheating, radiation, and matter eras.
Since the resulting analysis is very close to that performed in Ref.~\cite{Campanelli-Cea},
we quote just the final result (full details are given in Appendix~A):
\begin{equation}
\label{B0} \frac{B_0}{\nG} \sim \left(\frac{M}{10^{6} \GeV}
\right)^{\!\!6} \left(\frac{T_{\rm RH}}{\MeV} \right)^{\!\!7(n-1)/2}
\!\! \left(\frac{k_F}{10^{-37}} \right)^{\!\!n/2} \!\!,
\end{equation}
where $T_{\rm RH}$ is the reheat temperature (see Appendix~A),
and $n$ takes the values $\pm 1$ according to $T_* \lesssim T_{\rm
RH}$ or $T_* \gtrsim T_{\rm RH}$. The temperature $T_*$,
defined in Eq.~(\ref{Tstar}), is the temperature
below which electric fields are washed out by dissipative effects of the
primordial plasma (so that magnetic fields evolve adiabatically for $T \lesssim T_*$).
The condition $T_* \gtrsim T_{\rm RH}$ means, indeed,
that the magnetic field evolves adiabatically from the end
of reheating until today and is equivalent to having $T_{\rm RH}/\MeV
\lesssim (10^{-37}\!/k_F)^{1/7}$.

Upper bounds on $||(k_F)_{\alpha \beta \mu \nu}||$ come from
the analysis of CMB polarization and polarized light of
radiogalaxies and gamma-ray bursts: they are, respectively,
$10^{-30}$~\cite{Kostelecky-m1}, $10^{-32}$~\cite{Kostelecky-m2},
and $10^{-37}$~\cite{Kostelecky-m3}. (Although the former bound is
less stringent than the latter ones, it covers the whole portion of
coefficient space for Lorentz violation. The point-source nature of
radiogalaxies and gamma-ray bursts, instead, allows us to put
constraints only on limited portions of coefficient
space~\cite{Kostelecky-m1}.)

Now, it is clear from Eq.~(\ref{B0}) and Fig.~1 that, for a wide range
of values of parameters defining inflation (i.e., $M$ and $T_{\rm RH}$)
and Lorentz symmetry violation (i.e., $k_F$), the scaling-invariant, present-day magnetic
field can be as strong as $B_0 \sim \nG$, and then could naturally explain the
presence of galactic and extragalactic magnetic fields.


\begin{figure}[t]
\begin{center}
\includegraphics[clip,width=0.47\textwidth]{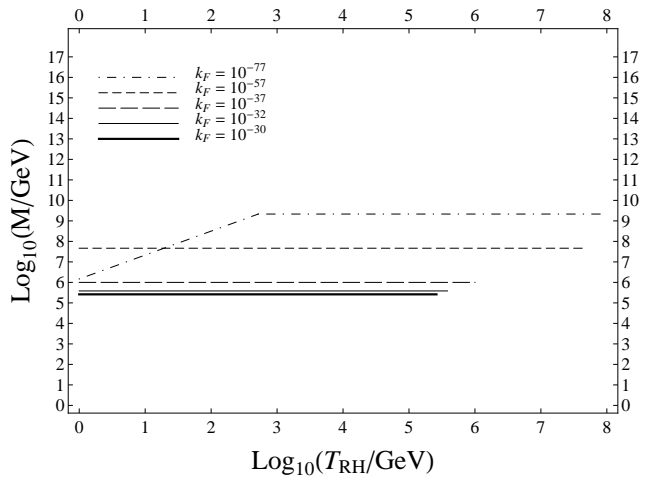}
\caption{The plane-symmetric, scale-invariant, inflation-produced
magnetic field has an actual intensity of order of nanogauss if the
values of parameter defining inflation, i.e. the energy scale of
inflation $M$ and the reheat temperature $T_{\rm RH}$, stay on the
curves. $k_F \sim ||(k_F)_{\alpha \beta \mu \nu}||$ estimates the
magnitude of the (constant) external tensor which parametrizes
Lorentz violation [see Eqs.~(\ref{Action}) and (\ref{k3003})].}
\end{center}
\end{figure}


\section{\normalsize{III. Ellipsoidal Universe and the CMB Quadrupole}}
\renewcommand{\thesection}{\arabic{section}}


The cosmic magnetic field produced according to the
mechanism discussed in the previous section possesses a planar
symmetry. The energy-momentum tensor for a
plane-symmetric magnetic field of the form $\B = (B_x,B_y,0)$ is
\begin{equation}
\label{E1} (T_{B}\!)^{\mu}_{\;\;\nu} =\!\!
    \left(
            \begin{array}{llll}
            \frac12 \B^2       & ~~~~~~~~\: 0                & ~~~~~~~~\: 0                 & ~~~~~ 0
            \\
            0                  & ~ \frac12 (B_x^2 - B_y^2)   & ~~~~~\, B_x B_y              & ~~~~~ 0
            \\
            0                  & ~~~~~\, B_x B_y             & ~ \frac12 (B_y^2 - B_x^2)    & ~~~~~ 0
            \\
            0                  & ~~~~~~~~\: 0                & ~~~~~~~~\: 0                 &  -\frac12 \B^2
            \end{array}
    \right) \!\! .
\end{equation}
Taking an ensemble average, and assuming statistical isotropy in the plane of symmetry, gives
$\langle B_x B_y \rangle \simeq 0$ and $\langle B_x^2 \rangle \simeq \langle B_y^2 \rangle$,
so that Eq.~(\ref{E1}) reduces to
%
%
$(T_{B})^{\mu}_{\;\;\nu} = \rho_B \, \mbox{diag} (1,0,0,-1)$,
where $\rho_B = \langle \B^2 \rangle/2$ is the average magnetic energy density.

We want now to analyze the large-scale evolution of the matter-dominated universe
filled with a plane-symmetric anisotropic component
represented by the planar cosmic magnetic field. The total energy-momentum tensor is then given by
\begin{equation}
\label{eq10bis}
T^{\mu}_{\;\;\nu} = \mbox{diag} (\rho_B + \rho_m,0,0,-\rho_B),
\end{equation}
where $\rho_m$ is the matter energy density.
We then consider a cosmological model with planar symmetry, whose
most general plane-symmetric line element, compatible with Eq.~(\ref{eq10bis}),
is~\cite{Taub}
\begin{equation}
\label{eq1}
ds^2 = dt^2 - a^2(t) (dx^2 + dy^2) - b^2(t) \, dz^2 ,
\end{equation}
where $a$ and $b$ are the scale factors. The metric (\ref{eq1})
corresponds to considering the $xy$-plane as a symmetry plane.

Taking into account Eqs.~(\ref{eq10bis}) and (\ref{eq1}),
Einstein's equations read~\cite{Berera}
\begin{eqnarray}
\label{eq5} && \left( \frac{\dot{a}}{a} \right)^{\!2} + 2 \,
\frac{\dot{a}}{a} \frac{\dot{b}}{b} = 8 \pi G (\rho_B+\rho_m), \\
\label{eq6} && \frac{\ddot{a}}{a} + \frac{\ddot{b}}{b} +
\frac{\dot{a}}{a} \frac{\dot{b}}{b} = 0, \\
\label{eq7} && 2 \, \frac{\ddot{a}}{a} + \left( \frac{\dot{a}}{a}
\right)^{\!2} = -8 \pi G \rho_B,
\end{eqnarray}
where a dot indicates the derivative with respect to the cosmic
time.

Since the conductivity of the primordial plasma is very high,
we are safe in neglecting interaction effects between magnetic field
and matter~\cite{Widrow}. In this case, the
magnetic component of the energy-momentum tensor is conserved,
$D_\mu (T_B)^{\mu}_{\;\; \nu} = 0$, so that we have
\begin{equation}
\label{cons_anisotr} {\dot{\rho}}_B + 2 \! \left( \frac{\dot{a}}{a} + \frac{\dot b}{b} \right) \! \rho_B  = 0.
\end{equation}
Let us introduce the so-called ``eccentricity''
\begin{equation}
\label{eccentr} e \equiv \sqrt{1 - \left( \frac{a}{b} \right)^{\!\!2}}
\, .
\end{equation}
The normalization of the scale factors is such that $a(t_0) =
b(t_0) = 1$ at the present time $t_0$.
\\
In this paper, we restrict our analysis to the case of small
eccentricities (that is, we consider the metric anisotropies as
perturbations over the isotropic Friedmann-Robertson-Walker
background). In this limit, from Eqs.~(\ref{eq5})-(\ref{eq7}), we
get the following evolution equation for the eccentricity:
\begin{equation}
\label{evolution} \frac{d (e \dot{e})}{dt} + 3 H (e \dot{e}) = 8
\pi G \rho_B.
\end{equation}
Here $H = \dot{a}/a$ is the usual Hubble expansion parameter for the
isotropic universe. In the matter-dominated era, it results in $a(t)
\propto t^{2/3}$, so that $H = 2/(3t)$. Moreover, to the zero-order
in the eccentricity, from Eq.~(\ref{cons_anisotr}) it follows that
the magnetic energy density scales in time as $\rho_B \propto
a^{-4}$. The solution of Eq.~(\ref{evolution}) is
\begin{equation}
\label{eq16} e^2 = 4 \Omega_B^{(0)} \! \left( 1 - \frac{3}{a} +
\frac{2}{a^{3/2}} \right) \!,
\end{equation}
%
%
where $\Omega_B^{(0)} = \rho_B(t_0)/\rho_{\rm cr}^{(0)}$,
$\rho_{\rm cr}^{(0)} = 3 H_0^2/8 \pi G$ is the actual critical
energy density, and $H_0 = 100 \, h \, {\rm km}\,{\rm
sec}^{-1}{\rm Mpc}^{-1}$ is the current Hubble parameter with $h
\simeq 0.72$ the little-$h$ constant~\cite{WMAP5}. At the
decoupling, $t=t_{\rm dec}$, we have $e_{\rm dec}^2 \simeq 8 \,
\Omega_B^{(0)} z_{\rm dec}^{3/2}$, where $e_{\rm dec} = e(t_{\rm
dec})$ and $z_{\rm dec} \simeq 1091$ is the red-shift at
decoupling~\cite{WMAP5}. Accordingly, we get
\begin{equation}
\label{eccentricity2} e_{\rm dec} \simeq 4 \times 10^{-3} \,
\frac{B_0}{\nG} \, .
\end{equation}
If, for instance, we assume for the present cosmological magnetic
field strength the estimate $B_0 \simeq \nG$, which is compatible
with the constraints derived in Ref.~\cite{Ferreira}, with the presence
of galactic magnetic fields~\cite{Widrow}, and with the results obtained in Sec. II,
we get an eccentricity at decoupling of about $e_{\rm dec} \simeq 0.4 \times
10^{-2}$.


The presence of a nanogauss, plane-symmetric magnetic field at the surface of last scattering
affects the temperature anisotropies in such a way, as we are going to see, to solve
the quadrupole problem.

Expanding the temperature anisotropy in terms of spherical
harmonics~\cite{Dodelson}
\begin{equation}
\label{DeltaT}
\frac{\Delta T(\theta,\phi)}{\langle T \rangle} =
\sum_{l=2}^{\infty} \sum_{m=-l}^{l} a_{lm} Y_{lm}(\theta,\phi),
\end{equation}
and introducing the power spectrum
\begin{equation}
\label{spectrum}
\frac{\Delta T_l}{\langle T \rangle} = \sqrt{ \frac{1}{2 \pi} \,
\frac{l(l+1)}{2l+1} \sum_m |a_{lm}|^2},
\end{equation}
the quadrupole anisotropy is defined by the multipole
$\ell=2$
\begin{equation}
\label{quadrupole-T}
\mathcal{Q} \, \equiv \, \frac{\Delta T_2}{\langle T \rangle} \, ,
\end{equation}
where $\langle T \rangle \simeq 2.73$K is the actual (average)
temperature of the CMB radiation. The quadrupole problem resides in
the fact that the observed quadrupole anisotropy, according to 3-year WMAP
data (see Table I) and 5-year WMAP data (see Table II), is in the range
\begin{equation}
\label{quad-obs} \left(\Delta T_2 \right)_{\rm obs}^{2} \simeq (210
\div 276) \, \mu \mbox{K}^2
\end{equation}
and
\begin{equation}
\label{quad-obs2} \left(\Delta T_2 \right)_{\rm obs}^{2} \simeq (213
\div 302) \, \mu \mbox{K}^2,
\end{equation}
respectively, while the expected quadrupole anisotropy according the
$\Lambda$CDM standard model is
\begin{equation}
\label{quad-infl} \left(\Delta T_2 \right)_{\rm I}^2 \simeq 1252 \,
\mu \mbox{K}^2,
\end{equation}
if we take into account the 3-year WMAP data, or
\begin{equation}
\label{quad-infl2} \left(\Delta T_2 \right)_{\rm I}^2 \simeq 1207 \,
\mu \mbox{K}^2,
\end{equation}
according to 5-year WMAP data.

\begin{table}[h!]
\begin{center}
\caption{The cleaned maps SILC400, WILC3YR, and TCM3YR. Note that
the values of $a_{2m}$ in this table correspond to the values of
$a_{2m}$ given in Refs.~\cite{SILC400,WILC3YR,TCM3YR} divided by
$\langle T \rangle \simeq 2.73 K$. Moreover, the values of $\mbox{Re}[a_{2m}]$
and $\mbox{Im}[a_{2m}]$ are in units of $10^{-6}$.} \vspace{0.5cm}
\begin{tabular}{lllllll}

\hline \hline

&Map     &$m$ &$\mbox{Re}[a_{2m}]$  &$\mbox{Im}[a_{2m}]$  &$\left(\Delta T_2\right)^2\!/\mu\mbox{K}^2$  &$\mathcal{Q}/10^{-6}$ \\
\hline
&        &$0$ &~~\:$2.75$           &~~\:$0.00$           &                                             & \\
&SILC400 &$1$ &~\!$-0.56$           &~~\:$1.77$           &~~~~\:$275.8$                                & \; $6.1$ \\
&        &$2$ &~\!$-6.79$           &~\!$-6.60$           &                                             & \\
\hline
&        &$0$ &~~\:$4.21$           &~~\:$0.00$           &                                             & \\
&WILC3YR &$1$ &~\!$-0.02$           &~~\:$1.78$           &~~~~\:$248.8$                                & \; $5.8$ \\
&        &$2$ &~\!$-5.28$           &~\!$-6.89$           &                                             & \\
\hline
&        &$0$ &~~\:$1.22$           &~~\:$0.00$           &                                             & \\
&TCM3YR  &$1$ &~~\:$0.10$           &~~\:$1.79$           &~~~~\:$209.5$                                & \; $5.3$ \\
&        &$2$ &~\!$-5.45$           &~\!$-6.32$           &                                             & \\

\hline \hline

\end{tabular}
\end{center}
\end{table}


\begin{table}[h!]
\begin{center}
\caption{Quadrupole power obtained from the cleaned maps ``Hinshaw
{\it et al.} cut sky'', WILC5YR, and HILCM5YR. Data from
Ref.~\cite{Kim}.} \vspace{0.5cm}
\begin{tabular}{lllllll}

\hline \hline

&Map                            &~~$\left(\Delta T_2\right)^2\!/\mu\mbox{K}^2$  &~~$\mathcal{Q}/10^{-6}$ \\
\hline

& Hinshaw {\it et al.} cut sky  &~~~~~~\;$213.4$ &~~~\,$5.4$  \\
& WILC5YR                       &~~~~~~\;$242.7$ &~~~\,$5.7$  \\
& HILCM5YR                      &~~~~~~\;$301.7$ &~~~\,$6.4$  \\

\hline \hline

\end{tabular}
\end{center}
\end{table}


If we admit that the large-scale spatial geometry of our Universe is
plane-symmetric with a small eccentricity, then we have that the
observed CMB anisotropy map is a linear superposition of two
contributions~\cite{prl,Bunn}:
%
%
$\Delta T \; = \; \Delta T_{\rm A} + \, \Delta
T_{\rm I}$,
where $\Delta T_{\rm A}$ represents the temperature fluctuations
due to the anisotropic spacetime background, while $\Delta T_{\rm
I}$ is the standard isotropic fluctuation caused by the
inflation-produced gravitational potential at the last scattering
surface. As a consequence, we may write
\begin{equation}
\label{alm} a_{lm} = a_{lm}^{\rm A} + \, a^{\rm I}_{lm}.
\end{equation}
We want now to analyze the distortion of the CMB radiation in a
universe with planar symmetry described by the metric (\ref{eq1})
in the small eccentricity approximation.
The null geodesic equation gives that a photon emitted at
the last scattering surface having energy $E_{\rm dec}$ reaches
the observer with an energy equal to
$E_0(\widehat{n}) = \langle E_0 \rangle (1 + e_{\rm dec}^2 n_3^2/2)$,
where $\langle E_0 \rangle \equiv E_{\rm dec}/(1+z_{\rm dec})$, and
$\widehat{n} = (n_1,n_2,n_3)$ are the direction cosines of the null
geodesic in the symmetric (Robertson-Walker) metric.

It is worth mentioning that the above result applies to the
special case where the normal to the plane of symmetry is
directed along the $z$-axis. The general case where this normal is
directed along an arbitrary direction in a coordinate system
$(x_{\rm G},y_{\rm G},z_{\rm G})$ in which the $x_{\rm G} y_{\rm G}$-plane
is the galactic plane, has been analyzed in Ref.~\cite{prl}.
Closely following~\cite{prl}, we perform a rotation
$R = R_x(\vartheta) \, R_z(\varphi + \pi/2)$
of the coordinate system $(x,y,z)$, where $R_z(\varphi + \pi/2)$ and
$R_x(\vartheta)$ are rotations of angles $\varphi + \pi/2$ and
$\vartheta$ about the $z$- and $x$-axis, respectively. In
the new coordinate system the $z$-axis is directed along the
direction defined by the polar angles $(\vartheta, \varphi)$.
Therefore, the temperature anisotropy in this new reference system is
\begin{equation}
\label{DeltaTA} \frac{\Delta T_{\rm A}}{\langle T \rangle} \equiv
\frac{E_0(n_{\rm A}) -\langle E_0 \rangle}{\langle E_0 \rangle} =
\frac{1}{2} \, e_{\rm dec}^2 n_{\rm A}^2,
\end{equation}
where $n_{\rm A} \equiv (R \, \widehat{n})_3$ is given by
\begin{equation}
\label{nA} n_{\rm A}(\theta,\phi) = \cos \theta \cos \vartheta -
\sin \theta \sin \vartheta \cos(\phi - \varphi).
\end{equation}
Equations (\ref{DeltaTA}) and (\ref{nA}), then, give the general expression
for the temperature anisotropy induced by a planar metric whose normal
to the plane of symmetry points in the direction
$(\vartheta, \varphi)$ in the galactic coordinate system.

From Eqs.~(\ref{DeltaTA}) and (\ref{nA}), it follows that only the
quadrupole terms ($\ell = 2$) are different from zero:
\begin{eqnarray}
\label{almA}
&& \!\!\!\!\!\!\!\! a_{20}^{\rm A} = \frac{\sqrt{\pi}}{6\sqrt{5}} \,
                  [1 + 3\cos(2 \vartheta) ] \, e_{\rm dec}^2 \, , \nonumber \\
&& \!\!\!\!\!\!\!\! a_{21}^{\rm A} = -(a_{2,-1}^{\rm A})^{*} =
                  -\sqrt{\frac{\pi}{30}} \;
                  e^{-i \varphi}  \sin(2\vartheta) \, e_{\rm dec}^2 \, , \\
&& \!\!\!\!\!\!\!\! a_{22}^{\rm A} = (a_{2,-2}^{\rm A})^{*} =
                  \sqrt{\frac{\pi}{30}}
                  \; e^{-2 i \varphi} \sin^2\!\vartheta \,
                  e_{\rm dec}^2 \, . \nonumber
\end{eqnarray}
%
%
%
Since the temperature anisotropy is a real function, we have
$a_{l,-m} = (-1)^m (a_{l,m})^*$. Observing that $a_{l,-m}^{\rm A}
= (-1)^m (a_{l,m}^{\rm A})^*$ [see Eq.~(\ref{almA})], we get
$a^{\rm I}_{l,-m} = (-1)^m (a^{\rm I}_{l,m})^*$.
Moreover, because the standard inflation-produced temperature
fluctuations are statistically isotropic, we will make the
reasonable assumption that the $a^{\rm I}_{2m}$ coefficients are
equals up to a phase factor. Therefore, we can write~\cite{prl}
\begin{eqnarray}
\label{almI}
&& a^{\rm I}_{20} = \sqrt{\frac{\pi}{3}} \; e^{i \phi_1} \mathcal{Q}_{\rm I}, \nonumber \\
&& a^{\rm I}_{21} =  - (a^{\rm I}_{2,-1})^{*} = \sqrt{\frac{\pi}{3}} \; e^{i \phi_2} \mathcal{Q}_{\rm I} \; , \\
&& a^{\rm I}_{22} =  (a^{\rm I}_{2,-2})^{*} = \sqrt{\frac{\pi}{3}}
\; e^{i \phi_3} \mathcal{Q}_{\rm I} \; , \nonumber
\end{eqnarray}
where $0 \leq \phi_1 \leq 2 \pi$, $0 \leq \phi_2 \leq 2 \pi$, and $0 \leq \phi_3 \leq 2 \pi$
are unknown phases, and
\begin{equation}
\label{QInflation} \mathcal{Q}_{\rm I} \simeq 13.0 \times 10^{-6}
\end{equation}
or
\begin{equation}
\label{QInflation2} \mathcal{Q}_{\rm I} \simeq 12.7 \times 10^{-6}
\end{equation}
according to the 3-year WMAP data or 5-year WMAP data, respectively.

%
%
%
%

We may fix the unknown direction $(\vartheta, \varphi)$ and
the eccentricity by solving Eq.~(\ref{alm}), which is (for
$\ell=2$) a system of 5 equations containing 5 unknown parameters:
$e_{\rm dec}$, $\vartheta$, $\varphi$, $\phi_2$, and $\phi_3$.
Note that it is always possible to choose $a_{20}$ real, so that $\phi_1 = 0$.


\begin{table}[h!]
\begin{center}
\caption{Numerical solutions of Eq.~(\ref{alm}) obtained by using
the map SILC400. The values of the angles $\vartheta$, $\varphi$,
$\phi_2$, and $\phi_3$ are in degrees.} \vspace{0.5cm}
\begin{tabular}{lllllll}

\hline \hline

&$e_{\dec}/10^{-2}$ &$~~~\vartheta$ &$~~~~~\varphi$ &$~~~~~\phi_2$ &$~~~~\,\phi_3$ &$~~~~B_0/\nG$ \\

\hline

&$~~0.76$           &$~67.9$        &$~~~5.8$       &$~~~31.4$     &$~~~18.7$      &$~~~~~~1.88$  \\
&$~~0.76$           &$~67.8$        &$~~~174.4$     &$~~~33.9$     &$~~~69.4$      &$~~~~~~1.89$  \\
&$~~0.76$           &$~67.8$        &$~~~354.3$     &$~~~121.2$    &$~~~69.8$      &$~~~~~~1.89$  \\
&$~~0.76$           &$~67.7$        &$~~~185.5$     &$~~~121.1$    &$~~~19.5$      &$~~~~~~1.89$  \\
&$~~0.76$           &$~67.9$        &$~~~50.1$      &$~~~121.1$    &$~~~157.9$     &$~~~~~~1.88$  \\
&$~~0.76$           &$~67.9$        &$~~~130.0$     &$~~~138.9$    &$~~~111.9$     &$~~~~~~1.88$  \\
&$~~0.76$           &$~67.7$        &$~~~310.3$     &$~~~131.9$    &$~~~112.1$     &$~~~~~~1.89$  \\
&$~~0.76$           &$~67.7$        &$~~~229.6$     &$~~~131.9$    &$~~~157.6$     &$~~~~~~1.90$  \\

\hline \hline

\end{tabular}
\end{center}
\end{table}


\begin{table}[h!]
\begin{center}
\caption{As in Table III, but for the map WILC3YR.} \vspace{0.5cm}
\begin{tabular}{lllllll}

\hline \hline

&$e_{\dec}/10^{-2}$ &$~~~\vartheta$ &$~~~~~\varphi$ &$~~~~~\phi_2$ &$~~~~\,\phi_3$ &$~~~~B_0/\nG$ \\

\hline

&$~~0.74$           &$~66.4$        &$~~~4.5$       &$~~~36.7$     &$~~~30.8$      &$~~~~~~1.84$  \\
&$~~0.74$           &$~66.3$        &$~~~175.5$     &$~~~36.8$     &$~~~72.3$      &$~~~~~~1.84$  \\
&$~~0.74$           &$~66.3$        &$~~~355.6$     &$~~~110.4$    &$~~~72.2$      &$~~~~~~1.84$  \\
&$~~0.74$           &$~66.3$        &$~~~184.4$     &$~~~110.4$    &$~~~31.1$      &$~~~~~~1.84$  \\
&$~~0.74$           &$~66.5$        &$~~~57.3$      &$~~~140.0$    &$~~~160.5$     &$~~~~~~1.83$  \\
&$~~0.74$           &$~66.5$        &$~~~122.7$     &$~~~140.0$    &$~~~112.0$     &$~~~~~~1.83$  \\
&$~~0.75$           &$~66.2$        &$~~~303.3$     &$~~~130.8$    &$~~~112.5$     &$~~~~~~1.85$  \\
&$~~0.75$           &$~66.2$        &$~~~236.7$     &$~~~130.8$    &$~~~160.0$     &$~~~~~~1.85$  \\

\hline \hline

\end{tabular}
\end{center}
\end{table}


\begin{table}[h!]
\begin{center}
\caption{As in Table III, but for the map TCM3YR.} \vspace{0.5cm}
\begin{tabular}{lllllll}

\hline \hline

&$e_{\dec}/10^{-2}$ &$~~~\vartheta$ &$~~~~~\varphi$ &$~~~~~\phi_2$ &$~~~~\,\phi_3$ &$~~~~B_0/\nG$ \\

\hline

&$~~0.78$           &$~69.2$        &$~~~0.5$       &$~~~89.3$     &$~~~46.2$      &$~~~~~~1.94$  \\
&$~~0.78$           &$~69.2$        &$~~~179.4$     &$~~~89.0$     &$~~~52.5$      &$~~~~~~1.94$  \\
&$~~0.78$           &$~69.2$        &$~~~359.5$     &$~~~96.6$     &$~~~52.3$      &$~~~~~~1.94$  \\
&$~~0.78$           &$~69.2$        &$~~~180.6$     &$~~~96.9$     &$~~~49.5$      &$~~~~~~1.94$  \\
&$~~0.78$           &$~69.3$        &$~~~49.0$      &$~~~140.5$    &$~~~154.5$     &$~~~~~~1.93$  \\
&$~~0.78$           &$~69.3$        &$~~~131.1$     &$~~~140.4$    &$~~~116.4$     &$~~~~~~1.93$  \\
&$~~0.78$           &$~69.1$        &$~~~311.6$     &$~~~130.6$    &$~~~116.6$     &$~~~~~~1.94$  \\
&$~~0.78$           &$~69.1$        &$~~~228.4$     &$~~~130.6$    &$~~~154.3$     &$~~~~~~1.94$  \\

\hline \hline

\end{tabular}
\end{center}
\end{table}


To solve Eq.~(\ref{alm}) for $\ell=2$, we need the observed values
of the $a^{\rm}_{2m}$'s. We use the cleaned CMB temperature
fluctuation map of the 3-year WMAP data obtained by using an improved
internal linear combination method as galactic foreground
subtraction technique. In particular, we adopt the three maps
SILC400~\cite{SILC400}, WILC3YR~\cite{WILC3YR}, and
TCM3YR~\cite{TCM3YR}. For completeness, we report in Table I the
values of $a^{\rm}_{2m}$ corresponding to these maps.

Numerical solutions of Eq.~(\ref{alm}), referring to the three maps,
are given in Tables III, IV, and V, respectively. In Appendix~B, we
will show that the system (\ref{alm}) (for $\ell=2$) admits 8 independent
solutions. Moreover we observe that, for each independent solution
$(e_{\rm dec},\vartheta,\varphi,\phi_2,\phi_3)$ shown in the tables,
there exists another one given by $(e_{\rm
dec},\pi-\vartheta,\varphi \pm \pi,\phi_2,\phi_3)$, where we take
the plus sign if $\varphi < \pi$ and the minus sign if $\varphi >
\pi$.

Looking at Table I and Eq.~(\ref{QInflation}), we see that the
values of coefficients $|a_{20}|$ and $|a_{21}|$ are much smaller
than $\mathcal{Q}_{\rm I}$. Moreover, comparing
Eq.~(\ref{QInflation}) with Table I, and Eq.~(\ref{QInflation2})
with Table II, we deduce, roughly speaking, that the value of
$\mathcal{Q}_{\rm I}$ is about twice the value of $\mathcal{Q}$.
Assuming $\mathcal{Q}_{\rm I} \gg |a_{20}|, |a_{21}|$, and
$\mathcal{Q}_{\rm I} \simeq 2 \mathcal{Q}$, we will show in
Appendix~B that approximate solutions for $e_{\rm dec}$, $\vartheta$,
and $\varphi$ are
\begin{eqnarray}
\label{eapprox} && e_{\rm dec}^2 \simeq \frac{\sqrt{15}
(5+3\sqrt{73})}{24} \, \mathcal{Q}_{\rm I},
\\
\label{tetaapprox} && \vartheta \simeq \frac{\pi}{2} - 
\frac{1}{2} \arcsin \! \left[ \frac{\sqrt{6} (3\sqrt{73}-5)}{79}
\right] \! ,
\\
\label{phiapprox} && \varphi \simeq 0, \frac{\pi}{4} \, ,
\frac{3\pi}{4} \, , \pi, \frac{5\pi}{4} \, ,\frac{7\pi}{4} \, ,
2\pi.
\end{eqnarray}
From Eqs.~(\ref{eapprox}) and (\ref{QInflation})-(\ref{QInflation2}) we get
\begin{equation}
\label{Result1} e_{\rm dec} \simeq 0.8 \times
10^{-2},
\end{equation}
%
which inserted in Eq.~(\ref{eccentricity2}) gives
\begin{equation}
\label{Result2} B_0 \simeq 2 \, \nG.
\end{equation}
%
As one can check, the approximate solutions
Eqs.~(\ref{eapprox}), (\ref{tetaapprox}), and (\ref{phiapprox}) are
quite close to the numerical values.



\begin{figure}[t]
\begin{center}
\includegraphics[clip,width=0.45\textwidth]{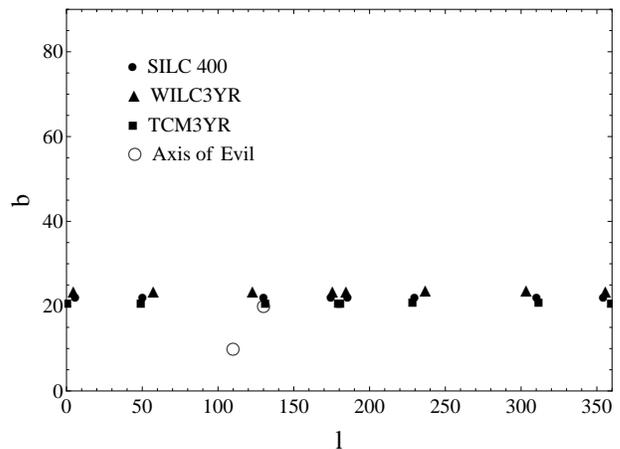}
\caption{Numerical solutions of Eq.~(\ref{alm}) obtained by using
the three maps in Table I. Note that $b = 90^{\circ} - \vartheta$,
and $l = \varphi$, where $(b,l)$ are the galactic coordinates. The
two open circles at $(b,l) \simeq (20^{\circ},130^{\circ})$ and
$(b,l) \simeq (10^{\circ},110^{\circ})$ define the direction of the
axis of evil determined in Ref.~\cite{Groeneboom} for the
5-year WMAP temperature sky maps in V-band and W-band, respectively.}
\end{center}
\end{figure}


In Fig.~2, we plot the numerical solutions $\vartheta$ versus
$\varphi$ which, defining a symmetry axis, singles out a preferred direction in the CMB map.
We adopt the so-called galactic coordinates system characterized by the
galactic latitude, $b$, and galactic longitude, $l$. In our
notation, the angle $b$ corresponds to $b = 90^{\circ} - \vartheta$
while $l = \varphi$. Equation~(\ref{tetaapprox}) gives
\begin{equation}
\label{Result3} b \simeq 20^{\circ}, 
\end{equation}
while Eq.~(\ref{phiapprox}) can be written as
\begin{equation}
\label{Result4} l \simeq 0^{\circ}, 45^{\circ}, 135^{\circ}, 180^{\circ}, 225^{\circ}, 315^{\circ}, 360^{\circ}.
\end{equation}
From the above equations and Fig.~2, we see that the galactic latitude
of the symmetry axis is remarkably independent on the adopted CMB
temperature fluctuation map, while galactic longitude is poorly constrained.

It is interesting to observe that, according to Ref.~\cite{Groeneboom},
the axis of evil points toward
\begin{equation}
\label{AE-V} (b,l)_{\rm AE} \simeq (20^{\circ},130^{\circ})
\end{equation}
if one takes into account the 5-year WMAP data in the V-band,
and toward
\begin{equation}
\label{AE-W} (b,l)_{\rm AE} \simeq (10^{\circ},110^{\circ})
\end{equation}
if one uses the W-band 5-year WMAP data, so that it seems to be
very close to the direction
\begin{equation}
(b,l) \simeq (20^{\circ},135^{\circ})
\end{equation}
defined by the symmetry axis in our model.

It is important to stress, however, that the apparent alignment of the above two axes
could be just an accidental fact. Indeed, the axis of evil is determined by a
statistical correlation (alignment and planarity) between the quadrupole and
octupole modes, while the presence of a planar cosmic magnetic field affects
only the quadrupole moment in the CMB radiation (this is true to the lowest
order in the eccentricity parameter considered in this paper; to higher orders,
modifications to the standard value of octupole intensity can occur, but they
are vanishingly small).
\\
In any case, it is worth noting that there are already independent
indications of a symmetry axis in the large-scale geometry of the
Universe as, for example, those coming from the analysis of spiral galaxies in the Sloan
Digital Sky Survey~\cite{Longo}.


\section{\normalsize{IV. Conclusions}}
\renewcommand{\thesection}{\arabic{section}}

Our knowledge of the Universe has greatly improved after the discovery
of the cosmic microwave background radiation. The observation and the
analysis of the temperature fluctuations in this relic radiation,
especially in the last decade, has confirmed at a surprising level of accuracy
the canonical theoretical model describing the evolution of the Universe,
the {\it standard cosmological model}. This is the celebrated hot big-bang
model, ``equipped'' with inflation, dark energy, and cold dark matter.
Although these last three ingredients have not yet been framed in
a definitely theoretical model of particle physics, from an operative point of view,
this cosmological landscape appears satisfactory.

However, the ubiquitous existence of large-scale correlated magnetic fields
and some presumed anomaly discovered in the spectrum of the CMB radiation,
together with the statement of fact that the evolution of the Universe (at least)
before the Planck era cannot be understood without a self-consistent theory
of quantum gravity, justify the study of both alternative cosmologies
(such as anisotropic models of the Universe) and models of particle physics beyond
the standard model.

Indeed, in this paper, we analyzed the effects on standard cosmology of including in the
photon sector of the standard model a Lorentz-violating term, the ``Kosteleck\'{y} term''
$\mathcal{L}_K = -\mbox{$\frac14$} (k_F)_{\alpha \beta \mu \nu} F^{\alpha \beta} F^{\mu \nu}$.
Remarkably, we found that its presence is responsible for a
superadiabatic amplification of magnetic vacuum fluctuations during de Sitter inflation.
Moreover, the amplified magnetic field possesses a planar symmetry at large
cosmological scales if the  external tensor $(k_F)_{\alpha \beta \mu \nu}$,
which parametrizes Lorentz violation, assumes a particular (asymmetric) form.
This peculiar magnetic field could account for the presence
galactic magnetic fields and induces a small anisotropization of the Universe
at cosmological scales (which can be described by a Bianchi I model).
This, in turn, could explain the low quadrupole anomaly in the CMB radiation.

Finally, we would like to stress that anisotropic cosmological models, such as that
induced by the planar cosmic magnetic field discussed in this paper,
are expected to induce a certain amount of polarization in
the CMB radiation~\cite{Negroponte}.
Although it is beyond the aim of this paper to discuss this kind
of problems, we notice that it has been shown~\cite{Cea} that the 3-year WMAP
data on large-scale polarization could be in agreement
with an anisotropic model of a universe of Bianchi I type.
An appropriate analysis of CMB polarization, resulting from the
cosmological model presented in this paper, is in progress.


\begin{acknowledgments}
\end{acknowledgments}



\section{\normalsize{Appendix A}}


In this appendix, we study the evolution during reheating, radiation and matter eras,
of the magnetic field produced, according to the mechanism discussed in Sec. II,
during de Sitter inflation.

After inflation, the Universe enters into the so-called reheating
phase, during which the energy of the inflaton is converted into
ordinary matter. The reheating phase ends at the temperature
$T_{\rm RH}$ which is less than $M$ and constrained as $T_{\rm RH}
\lesssim 10^8 \GeV$~\cite{Riotto}. Moreover, CMB analysis requires
$M \lesssim 10^{-2}\mPl$~\cite{Turner}, otherwise it would be too
much of a gravitational waves relic abundance, and also one must
impose that $T_{\rm RH} \gtrsim 1\GeV$, so that the predictions of
big-bang nucleosynthesis (BBN) are not spoiled~\cite{Turner,Sarkar}.

It is worth noting that the condition $k_F \gg a^6$ is certainly
fulfilled during inflation and reheating if $k_F \gg a_{\rm RH}^6$,
where $a_{\rm RH} = a(T_{\rm RH})$. Since $a_{\rm RH} \sim
T_0/T_{\rm RH}$, where $T_0 \sim 10^{-13} \GeV$ is the actual
temperature~\cite{Kolb}, we have $10^{-126} \lesssim a_{\rm RH}^6
\lesssim 10^{-78}$ for $1\GeV \lesssim T_{\rm RH} \lesssim 10^8
\GeV$. Taking into account that the most stringent bound on
$||(k_F)_{\alpha \beta \mu \nu}||$ is $10^{-37}$ (as discussed in Sec. II)
we get, for a wide range of allowed
values of $k_F$, that $k_F\gg a^6$ during inflation and reheating.
This, in turn means, according to Eqs.~(\ref{evolution2}), (\ref{k3003})
and the first Bianchi identity,
that a superadiabatic amplification of magnetic
vacuum fluctuations during these eras takes place. During reheating, in particular,
taking into account that $\eta \propto a^{1/2}$, we have $B \propto a^{9/2}$.


After reheating, the Universe enters the radiation-dominated era. In
this era, as well as in the subsequent matter era, the effects of
the conducting primordial plasma are important when studying the
evolution of a magnetic field. They are taken into account by adding
to the electromagnetic Lagrangian the source term $j^{\mu} \!
A_{\mu}$~\cite{Turner}. Here, the external current $j^{\mu}$,
expressed in terms of the electric field, has the form $j^{\mu} =
(0, \sigma_c {\textbf E})$, where $\sigma_c$ is the conductivity.
Plasma effects introduce, in the right-hand side of
Eq.~(\ref{motion}), the extra term $-a\sigma_c (a^2 E_i)$. In this
case, it easy to see that modes well outside the horizon (assuming
that $k_F \gg a^6$) evolve as
\begin{equation}
E \propto a^4 \exp \! \left( \! -\frac{1}{2k_F} \int \!d\eta \, a^7\sigma_c \! \right) \! .
\end{equation}
Approximating $\int \!d\eta \, a^7\sigma_c$ with
$\eta \, a^7\sigma_c$ and using $a\eta \sim H^{-1}$, we get
\begin{equation}
E \propto a^4 \exp \! \left( \! -\frac{a^6\sigma_c}{2k_F H} \right) \!.
\end{equation}
In radiation era $H \sim T^2/\mPl$~\cite{Kolb} and,
for temperature much greater than the electron mass,
the conductivity is approximately given by
$\sigma_c \sim T/\alpha$~\cite{Turner}, where $\alpha$ is the fine
structure constant and $T$ the temperature. Then we get
\begin{equation}
E \propto a^4 \exp \!\! \left[ -\left( \frac{T_*}{T} \right)^{\!\!7} \, \right] \!,
\end{equation}
where
\begin{equation}
\label{Tstar} \frac{T_*}{\GeV} \sim \left( \frac{10^{-57}}{k_F} \right)^{\!1/7} \! .
\end{equation}
This means that for $T \gtrsim T_*$ we have
$E \propto a^4$ (which in turn gives $B \propto a^5$ since $\eta
\propto a$ in radiation era), while for $T \lesssim T_*$ the
electric field is dissipated, so the magnetic field evolves
adiabatically, $B \propto a^{-2}$. [We have assumed that $k_F \gg
a^6$ from the end of reheating until $T_*$. As is easy to verify
taking into account Eq.~(\ref{Tstar}), this assumption is certainly
satisfied in our case.]


Finally, evolving along the lines discussed above the
inflation-produced magnetic field from the time of horizon crossing
until today, and taking into account Eq.~(\ref{B1}),
we easily recover Eq.~(\ref{B0}).


\section{\normalsize{Appendix B}}

In this appendix, we solve the system of Eqs.~(\ref{alm}) 
for $\ell=2$. The numerical values of the $a_{2m}$'s
are listed in Table I, while the $a_{2m}^{\rm A}$'s and $a^{\rm
I}_{2m}$'s are given by Eqs.~(\ref{almA}) and (\ref{almI}),
respectively.

Since it is always possible to choose $a_{20}$ real, we take
$\phi_1 = 0$. Moreover, the temperature anisotropy is real so that
we have $a_{2,-m} = (-1)^m (a_{2m})^*$, $a_{2,-m}^{\rm A} = (-1)^m
(a_{2m}^{\rm A})^*$, and $a^{\rm I}_{2,-m} = (-1)^m (a^{\rm
I}_{2m})^*$.
Therefore, the system of Eqs.~(\ref{alm}) reduces to
\begin{eqnarray}
\label{system1}
&& a_{20} = \frac{\sqrt{\pi}}{6\sqrt{5}} \, [1 +
            3\cos(2\vartheta) ] \,
            e_{\rm dec}^2 \, + \,
            \sqrt{\frac{\pi}{3}} \, \mathcal{Q}_{\rm I}, \\
\label{system2}
&& \mbox{Re}[a_{21}] = -\sqrt{\frac{\pi}{30}} \,
                       \cos\!\varphi \, \sin(2\vartheta) \, e_{\rm dec}^2 \, + \,
                       \sqrt{\frac{\pi}{3}} \, \cos\!\phi_2 \mathcal{Q}_{\rm I}, \nonumber \\ \\
\label{system3}
&& \mbox{Im}[a_{21}] = \sqrt{\frac{\pi}{30}} \,
                       \sin\!\varphi \, \sin(2\vartheta) \, e_{\rm dec}^2 \, + \,
                       \sqrt{\frac{\pi}{3}} \, \sin\!\phi_2 \mathcal{Q}_{\rm I}, \nonumber \\ \\
\label{system4}
&& \mbox{Re}[a_{22}] = \sqrt{\frac{\pi}{30}} \,
                       \cos\!\varphi \, \sin^2\!\vartheta \, e_{\rm dec}^2 \, + \,
                       \sqrt{\frac{\pi}{3}} \, \cos\!\phi_3 \mathcal{Q}_{\rm I}, \nonumber \\ \\
\label{system5}
&& \mbox{Im}[a_{22}] = -\sqrt{\frac{\pi}{30}} \,
                       \sin\!\varphi \, \sin^2\!\vartheta \, e_{\rm dec}^2 \, + \,
                       \sqrt{\frac{\pi}{3}} \, \sin\!\phi_3 \mathcal{Q}_{\rm I}, \nonumber \\
\end{eqnarray}
where $\mathcal{Q}_{\rm I}$ is given by Eq.~(\ref{QInflation}). We
see that these  equations form a system of 5 transcendental
equations containing 5 unknown parameters: $e_{\rm dec}$,
$\vartheta$, $\varphi$, $\phi_2$, and $\phi_3$. Solving
Eq.~(\ref{system1}) with respect to $\vartheta$ we get two
independent solutions:
\begin{equation}
\label{tetatilde0} \vartheta = \{ \widetilde{\vartheta}, \pi
-\widetilde{\vartheta} \},
\end{equation}
where
\begin{equation}
\label{tetatilde} \widetilde{\vartheta} = \frac{1}{2} \arccos \!
\left( \frac{5\sqrt{15} \, a_{20} - 5\sqrt{5\pi}\mathcal{Q}_{\rm I}
- \sqrt{3\pi} \, e_{\rm dec}^2}{3\sqrt{3\pi} \, e_{\rm dec}^2}
\right) \! .
\end{equation}
Squaring Eqs.~(\ref{system4}) and (\ref{system5}), adding side by
side, and then solving with respect to $\varphi$, we obtain 8
independent solutions:
\begin{equation}
\label{phitilde0} \varphi = \{ \widetilde{\varphi}_{\pm}, \pi
-\widetilde{\varphi}_{\pm}, 2\pi -\widetilde{\varphi}_{\pm}, \pi +
\widetilde{\varphi}_{\pm} \},
\end{equation}
where
\begin{equation}
\label{phitilde} \widetilde{\varphi}_{\pm} = \frac{1}{2} \arccos \!
\left( \frac{\alpha \gamma \pm \beta \sqrt{\alpha^2 + \beta^2 -
\gamma^2}}{\alpha^2 + \beta^2} \right) \!,
\end{equation}
and
\begin{eqnarray}
\label{abc} \alpha \!\!&=&\!\!
-\sqrt{\frac{2\pi}{15}} \: \mbox{Re}[a_{22}] \sin^2\!\vartheta  \, e_{\rm dec}^2 \, , \\
\beta \!\!&=&\!\!
-\sqrt{\frac{2\pi}{15}} \: \mbox{Im}[a_{22}] \sin^2\!\vartheta \, e_{\rm dec}^2 \, , \\
\gamma \!\!&=&\!\! \frac{\pi}{30} \, \sin^4 \!\vartheta \, e_{\rm
dec}^4 + |a_{22}|^2 - \frac{\pi}{3} \mathcal{Q}_{\rm I}^2.
\end{eqnarray}
By dividing side by side Eqs.~(\ref{system3}) and (\ref{system2}),
and solving with respect to $\phi_2$, we get
\begin{equation}
\label{phi2} \tan\!\phi_2 = \frac{\sqrt{30} \, \mbox{Im}[a_{21}] -
\sqrt{\pi} \sin\!\varphi \, \sin(2\vartheta) \, e_{\rm
dec}^2}{\sqrt{30} \, \mbox{Re}[a_{21}] + \sqrt{\pi} \sin\!\varphi \,
\sin(2\vartheta) \, e_{\rm dec}^2} \, .
\end{equation}
The same procedure applied to Eqs.~(\ref{system5}) and
(\ref{system3}) results in
\begin{equation}
\label{phi3} \tan\!\phi_3 = \frac{\sqrt{30} \, \mbox{Im}[a_{22}] +
\sqrt{\pi} \sin(2\varphi) \, \sin^2\!\vartheta \, e_{\rm
dec}^2}{\sqrt{30} \, \mbox{Re}[a_{22}] - \sqrt{\pi} \sin(2\varphi)
\, \sin^2\!\vartheta \, e_{\rm dec}^2} \, .
\end{equation}
Finally, by squaring Eqs.~(\ref{system2}) and (\ref{system3}), and
adding side by side, we get
\begin{equation}
\label{eequation} e_{\rm dec}^4 - 2c \, e_{\rm dec}^2 - d = 0,
\end{equation}
where we have defined
\begin{eqnarray}
\label{cd1} c(\varphi,\vartheta) \!\!& = &\!\! \sqrt{\frac{30}{\pi}}
\,
(\mbox{Re}[a_{21}] \cos\varphi - \mbox{Im}[a_{21}] \sin\varphi) \csc(2\vartheta), \nonumber \\ \\
\label{cd2} d(\varphi,\vartheta) \!\!& = &\!\! 10 \! \left(
\mathcal{Q}_{\rm I}^2 - \frac{3}{\pi} \, |a_{21}|^2  \right) \!
\csc^2\!(2\vartheta).
\end{eqnarray}
We observe that the couple $(\vartheta, \varphi)$ can assume 16
different values, according to Eqs.~(\ref{tetatilde0}) and
(\ref{phitilde0}). Inserting these values in
Eqs.~(\ref{eequation})-(\ref{cd2}) we arrive at 16 different
equations for $e_{\rm dec}$. It is straightforward to verify that
only 8 of these are ``independent,'' in the sense that, given a
solution $(e_{\rm dec},\vartheta,\varphi)$ of one of the
independent equations, then $(e_{\rm dec}, \pi-\vartheta,
\varphi \pm \pi)$  is a solutions of one of the ``dependent'' ones
(we must  take the plus sign if $\varphi < \pi$ and the minus sign
if $\varphi > \pi$). The 8 independent equations can be solved
numerically and their solutions are presented in Table III, IV, and
V. \\
\indent
Finally, we derive the approximate solutions
(\ref{eapprox})-(\ref{tetaapprox}). To this end, we may formally
solve Eq.~(\ref{eequation}) to get $e_{\rm dec}^2 = c \pm
\sqrt{c^2 + d}$.
If $\mathcal{Q}_{\rm I} \gg |a_{21}|$ then $d \gg c^2$, and we
obtain
\begin{equation}
\label{eequation2} e_{\rm dec}^2 \simeq \sqrt{10} \,
\mathcal{Q}_{\rm I}\, |\csc(2\vartheta)|.
\end{equation}
Inserting the above expression in
Eqs.~(\ref{tetatilde0})-(\ref{tetatilde}), assuming that
$\mathcal{Q}_{\rm I} \gg |a_{20}|$, and after same algebraic
manipulation, we get
\begin{equation}
\label{sin2theta} |\sin(2\vartheta)| \simeq
\frac{\sqrt{6}(3\sqrt{73}-5)}{79} \, .
\end{equation}
Inserting the above equation in Eq.~(\ref{eequation2}) gives
Eq.~(\ref{eapprox}), while its solution is indeed
Eq.~(\ref{tetaapprox}).

Assuming that the main contribution to $\mathcal{Q}$ comes from
the $a_{22}$, and that $\mbox{Re}[a_{22}] \simeq
\mbox{Im}[a_{22}]$, we can write
\begin{equation}
\label{ReIm} \mbox{Re}[a_{22}] \simeq \mbox{Im}[a_{22}] \simeq -
\sqrt{\frac{5\pi}{12}} \, \mathcal{Q}.
\end{equation}
Consequently, inserting Eqs.~(\ref{eequation2})-(\ref{ReIm}) in
Eq.~(\ref{phitilde}), we get
\begin{equation}
\widetilde{\varphi}_{\pm} \simeq \frac12 \arccos \! \left( q \pm
\sqrt{\frac12 - q^2} \, \right) \! ,
\end{equation}
where
\begin{eqnarray}
q \!\!& = &\!\! \frac{\sqrt{30} (\sqrt{73} - 5)}{5760} \left[ (1 + 5\sqrt{73})
\frac{\mathcal{Q}_{\rm I}}{\mathcal{Q}} + 120
\frac{\mathcal{Q}}{\mathcal{Q}_{\rm I}} \right] \nonumber \\
\!\!& \simeq &\!\! 0.15 \frac{\mathcal{Q}_{\rm I}}{\mathcal{Q}} + 0.40
\frac{\mathcal{Q}}{\mathcal{Q}_{\rm I}} \, .
\end{eqnarray}
Observing that the value of the quadrupole according to the
$\Lambda$CDM standard model is, approximately, twice the value of
the observed one, $\mathcal{Q}_{\rm I} \simeq 2 \mathcal{Q}$
[compare Eq.~(\ref{QInflation}) with Table I, and
Eq.~(\ref{QInflation2}) with Table II], we roughly get $q \simeq 1/2$
and, accordingly, $\widetilde{\varphi}_{+} \simeq 0$ and
$\widetilde{\varphi}_{-} \simeq \pi/4$. Inserting these values in
Eq.~(\ref{phitilde0}), we recover Eq.~(\ref{phiapprox}).



\end{document}